\documentclass[12pt]{article}
\usepackage{epsf}
\begin{document}

\title{Semiclassical Approach to Parametric Spectral 
Correlation with Spin 1/2}

\author{Taro Nagao$^1$ and Keiji Saito$^2$}

\date{}
\maketitle

\begin{center}
\it
$^1$ Graduate School of Mathematics,
Nagoya University, Chikusa-ku, \\ Nagoya 464-8602, Japan \\
\it
$^2$ Department of Physics, Graduate School of Science,
University of Tokyo, Hongo 7-3-1, Bunkyo-ku, Tokyo 113-0033, Japan
\end{center}

\begin{abstract}

The spectral correlation of a chaotic system with spin $1/2$ 
is universally described by the GSE (Gaussian Symplectic 
Ensemble) of random matrices in the semiclassical limit. In 
semiclassical theory, the spectral form factor is expressed 
in terms of the periodic orbits and the spin state is simulated 
by the uniform distribution on a sphere. In this paper, instead 
of the uniform distribution, we introduce Brownian motion on 
a sphere to yield the parametric motion of the energy levels. 
As a result, the small time expansion of the form factor is 
obtained and found to be in agreement with the prediction of 
parametric random matrices in the transition within the GSE 
universality class. Moreover, by starting the Brownian motion 
from a point distribution on the sphere, we gradually increase 
the effect of the spin and calculate the form factor describing 
the transition from the GOE (Gaussian Orthogonal Ensemble) 
class to the GSE class.  

\end{abstract}


\medskip


\newpage

\section{Introduction}
\setcounter{equation}{0}
\renewcommand{\theequation}{1.\arabic{equation}}

The universal spectral correlation is one of the most 
outstanding features of quantum systems when the 
underlying classical dynamics is chaotic\cite{BGS}. It is known 
that there are universality classes depending on 
the symmetry of the systems. For example, 
if time reversal invariance is broken, the corresponding 
spectral correlation is reproduced by the GUE (Gaussian Unitary 
Ensemble) of random matrices. On the other hand, 
the spectral correlation of the systems with time 
reversal invariance depends on the spin. If the system 
is spinless or has an integer spin, the GOE (Gaussian 
Orthogonal Ensemble) gives a precise prediction, 
while the GSE (Gaussian Symplectic Ensemble)   
applies to a system with a half odd spin.       
\par
In order to explain the universal behaviour from the 
underlying chaotic dynamics, much effort has been paid 
to establish a semiclassical theory of spectral correlations. 
The spectral form factor $K(\tau)$ (the Fourier transform 
of the spectral correlation function) is one of the most 
typical quantities of interest. Berry first succeeded in 
evaluating the leading term in the semiclassical $\tau$ 
expansion of the spectral form factor\cite{BERRY}. Then 
Sieber and Richter specified the 
classical orbit pairs which contribute to the second 
order term\cite{SR}. More recently Heusler et al. and M\"uller et al. 
extended Sieber and Richter's work and calculate the full 
form of $K(\tau)$ in agreement with the prediction of 
random matrices\cite{HMBH04,MHBHA04,MHBHA05,SM,HMABH}.     
\par 
In addition to each of the universality classes, the 
transitions within and among them are also of interest. 
The transitions are described by the spectral correlations 
depending on the transition parameters. It is conjectured 
that such parametric correlations are also reproduced by 
parametric extensions of random matrices\cite{LH,HAAKE}. 
For the crossover from the GOE class to the GUE class, 
Saito and Nagao invented a scheme to incorporate the 
transition parameters into the semiclassical expansion 
of $K(\tau)$\cite{SN}. Similar schemes can also be 
applied to the transitions within the 
GUE and GOE classes\cite{PC,KS}. The agreements 
with parametric random matrices were in all cases confirmed.        
\par
In this paper, the parametric transition within the GSE 
symmetry class is treated. For that purpose, we shall 
study the spectral correlation of a chaotic system with 
spin $1/2$ by employing the strength of the effective field 
applied to the spin as the parameter. In order to simulate 
the spin dynamics, Brownian motion on the surface of 
a sphere is introduced. Using semiclassical periodic orbit 
theory, we evaluate the $\tau$ expansion of the spectral form 
factor up to the third order, so that the agreement with 
random matrix theory is confirmed. Moreover we study the 
crossover between a spinless system and a system with 
spin $1/2$. We suppose that the Brownian motion starts 
from a point distribution and that a diffusion on the sphere 
is caused by the increase of the coupling to the 
effective field. As a result, the semiclassical method 
yields the $\tau$ expansion of the form factor up to the 
second order. 
\par
The organization of this paper is as follows. In \S 2, 
semiclassical theory of a chaotic system with spin $1/2$ is 
developed. Assuming that the spin is coupled to a stochastic 
field, we explain how Brownian motion on a sphere 
arises. Then the leading term in the $\tau$ expansion of the form 
factor is evaluated by using Berry's diagonal approximation. 
In \S 3, a diagrammatic method is introduced to calculate 
the higher order terms in the $\tau$ expansion. In \S 4, 
the prediction of random matrix theory is presented and 
compared with the semiclassical result. In \S 5, a similar 
semiclassical analysis is carried out for the crossover 
from a spinless system to a system with spin $1/2$. The 
last section is devoted to a brief summary.  

\section{Periodic Orbit Theory for a Chaotic System with Spin 1/2}
\setcounter{equation}{0}
\renewcommand{\theequation}{2.\arabic{equation}}

Let us consider the energy level statistics of a bounded quantum system 
with $f$ degrees of freedom. Each phase space point is specified by a 
vector ${\bf x} = ({\bf q},{\bf p})$, where $f$ dimensional vectors 
${\bf q}$ and ${\bf p}$ give the position and momentum, respectively. 
It is assumed that the corresponding classical dynamics is chaotic (
homegeneously hyperbolic and ergodic). Moreover we suppose 
that the system has a spin with a fixed quantum number $S$. The 
strength of the interaction between the spin and effective field 
is characterized by a parameter $\eta$. 
\par
Let us denote by $E$ the energy of the system. Then, in the semiclassical 
limit $\hbar \to 0$, the energy level density $\rho(E;\eta)$ can be 
written in a decomposed form  
\begin{equation}
\rho(E;\eta) \sim \rho_{\rm av}(E) + \rho_{\rm osc}(E;\eta).
\end{equation}
Here $\rho_{\rm av}(E)$ is the local average of the level density, 
while $\rho_{\rm osc}(E;\eta)$ gives the fluctuation (oscillation) 
around the local average. 
\par
The local average of the level density is proportional to the 
number of Planck cells inside the energy shell:
\begin{equation}
\rho_{\rm av}(E) = (2 S + 1) \frac{\Omega(E)}{(2 \pi \hbar)^f},
\end{equation}
where the phase space volume with the energy between $E$ and $E + \Delta E$ 
is $\Omega(E) \Delta E$. The effective field is assumed to be so 
weak that $\rho_{\rm av}(E)$ does not depend on the parameter 
$\eta$.  
\par
On the other hand, in order to calculate the fluctuation part 
$\rho_{\rm osc}(E;\eta)$, we need to care about the time evolution 
of the spin. The spin state is described by a spinor with 
$2S + 1$ elements and the spin evolution operator ${\hat \Delta}$ 
is represented by a $(2 S + 1) \times (2 S + 1)$ matrix.  
We denote such a representation matrix evaluated along 
the periodic orbit $\gamma$ by $\Delta_{\gamma}(\eta)$. 
Then, in the leading order of the semiclassical approximation, the 
fluctuation part of the level density is written as\cite{MHBHA05,BK,KEPPELER}
\begin{equation}
\label{osc}
\rho_{\rm osc}(E;\eta) = \frac{1}{\pi \hbar} {\rm Re} \sum_{\gamma}
({\rm tr} \Delta_{\gamma}(\eta)) A_{\gamma} {\rm e}^{i S_{\gamma}(E)/\hbar}.
\end{equation}
Here $S_{\gamma}$ is the classical action for the orbital 
motion, $A_{\gamma}$ is the stability amplitude (including 
the Maslov phase) and ${\rm tr} \Delta_{\gamma}(\eta)$ is the 
sum of the diagonal elements of $\Delta_{\gamma}(\eta)$. 
\par   
Now we define the scaled parametric correlation function 
of the energy levels as
\begin{eqnarray}
R(s;\eta,\eta') & = & \left\langle \frac{
\rho\left(E + \frac{s}{2 \rho_{\rm av}(E)};\eta \right)
\rho\left(E - \frac{s}{2 \rho_{\rm av}(E)};\eta' \right)}{
\rho_{\rm av}(E)^2} \right\rangle - 1 \nonumber \\
& \sim & \left\langle \frac{
\rho_{\rm osc}\left(E + \frac{s}{2 \rho_{\rm av}(E)};\eta \right)
\rho_{\rm osc}\left(E - \frac{s}{2 \rho_{\rm av}(E)};\eta' \right)}{
\rho_{\rm av}(E)^2} \right\rangle.
\end{eqnarray}
Here we introduced averages depicted by the angular brackets 
$\langle \cdot \rangle$ over windows of the center energy $E$ 
and the scaled energy difference $s$. The form factor, namely the 
Fourier transform of $R(s;\eta,\eta')$, is then written as
\begin{eqnarray}
\label{ff}
K(\tau;\eta,\eta') & = &  
\int_{-\infty}^{\infty} {\rm d}s \ {\rm e}^{i 2 \pi \tau s}
R(s;\eta,\eta') \nonumber \\
& \sim & \left\langle \int {\rm d}\epsilon \ {\rm e}^{i \epsilon \tau
T_H/\hbar}
\frac{\rho_{\rm osc}\left(E + \frac{\epsilon}{2};\eta \right)
\rho_{\rm osc}\left(E - \frac{\epsilon}{2};\eta' \right)}{
\rho_{\rm av}(E)} \right\rangle.
\end{eqnarray}
Here the angular brackets mean averages over windows of 
the center energy $E$ and the time variable $\tau$. Note 
that $\tau$ is measured in units of the Heisenberg time
\begin{equation}
T_H = 2 \pi \hbar \rho_{\rm av}(E) = (2 S + 1) 
\frac{\Omega(E)}{(2 \pi \hbar)^{f-1}}.
\end{equation}
It follows from (\ref{osc}) and (\ref{ff}) that the form factor 
is expressed as a double sum over periodic orbits
\begin{equation}
\label{dsum}
K(\tau;\eta,\eta') 
\sim \frac{1}{T_H^2} \left\langle \sum_{\gamma,\gamma'}
({\rm tr}\Delta_{\gamma}(\eta))({\rm tr}\Delta_{\gamma'}(\eta'))^* 
A_{\gamma} A_{\gamma'}^* {\rm e}^{i (S_{\gamma} - S_{\gamma'})/
\hbar} \delta\left( \tau - \frac{T_{\gamma} + T_{\gamma'}}{2 T_H}
\right)
\right\rangle,
\end{equation}
where an asterisk stands for a complex conjugate. The periods of 
the periodic orbit $\gamma$ and its partner $\gamma'$ are denoted by 
$T_{\gamma}$ and $T_{\gamma'}$, respectively. 
\par
In principle, the spin evolution matrix $\Delta_{\gamma}(\eta)$ can be 
calculated from a deterministic equation of motion, if the Hamiltonian 
of the spin is explicitly known. However, here we take a simplified 
strategy based on an assumption that the spin evolution parameters 
undergo Brownian motion on the surface of a sphere\cite{HC}. The Brownian 
motion arises when the spin dynamics is determined by a stochastic Hamiltonian
\begin{equation}
\label{ham}
{\hat {\cal H}} = \eta ({\bf h} \cdot {\hat S}),
\end{equation}
where $\eta$ is an interaction-strength parameter and ${\hat S}$ 
is the spin operator. We assume that the components of the effective 
field 
\begin{equation}
{\bf h} = (h_x(t),h_y(t),h_z(t)) 
\end{equation}
can be replaced by isotropic Gaussian white noises: 
denoting the average over the noises by the brackets 
$\langle \langle \cdot \rangle \rangle$, 
we find the correlation
\begin{eqnarray}
& & \langle \langle h_j(t) h_l(t') \rangle \rangle = 0, \ \ \ j \neq l,  
\nonumber \\ 
& & \langle \langle h_j(t) h_j(t') \rangle \rangle = 2 D \delta(t - t') 
\end{eqnarray}
for $j,l = x,y,z$. Here isotropy implies that the diffusion 
constant $D$ does not depend on $j$. 
\par
The time evolution of the spin is described by a $(2 S + 1) 
\times (2 S + 1)$ matrix $\Delta(t)$ which satisfies the 
Schr\"odinger equation
\begin{equation}
\label{sch}
i \hbar \frac{\partial}{\partial t} \Delta(t) = {\cal H} 
\Delta(t),
\end{equation}
where ${\cal H}$ is the matrix representation of the Hamiltonian 
${\hat {\cal H}}$. Note that $\Delta(t)$ can be expressed as 
\begin{equation}
\label{delta}
\Delta(t) = {\rm exp}\left(i \phi(t) S_z/\hbar \right) 
 {\rm exp}\left(i \theta(t) S_x/\hbar \right) 
 {\rm exp}\left(i \psi(t) S_z/\hbar  \right),
\end{equation}
where $S_x$ and $S_z$ are $(2 S + 1) \times (2 S + 1)$ matrices 
representing the $x$ and $z$ components of the spin operator 
${\hat S}$. Thus three Euler angles $\psi$, $\theta$ and $\phi$ 
describe the spin evolution.  Let us denote by $\chi({\cal T})$ 
a segment (with the duration ${\cal T}$) of the periodic orbit 
$\gamma$. When ${\cal T}$ coincides with the period, $\chi({\cal T})$ 
is equated with $\gamma$. Along such a segment $\chi({\cal T})$, 
the spin evolution matrix $\Delta_{\chi({\cal T})}$ is evaluated as 
\begin{equation}
\label{ev}
\Delta_{\chi({\cal T})} = \Delta({\cal T}).
\end{equation} 
\par
Putting (\ref{delta}) into (\ref{sch}), we obtain the 
Langevin equation for the Euler angles 
\begin{eqnarray}
{\dot \phi}/\eta & = & h_x \sin\phi \cot \theta + h_y \cos \phi \cot \theta 
- h_z, \nonumber \\ 
{\dot \theta}/\eta & = & - h_x \cos \phi + h_y \sin \phi, \nonumber \\ 
{\dot \psi}/\eta & = & - h_x \sin\phi/\sin\theta - h_y \cos 
\phi/\sin\theta.
\end{eqnarray}
Then the Fokker-Planck equation 
\begin{equation}
\frac{\partial P}{\partial t} = \eta^2 D {\cal L}_{\rm SP} P
\end{equation}
holds for the p.d.f.(probability distribution function) $P(\psi,\theta,\phi)$ 
with the measure $\sin\theta {\rm d}\psi {\rm d}\theta {\rm d}\phi$. Here
\begin{equation}
{\cal L}_{\rm SP} = \frac{1}{\sin \theta} \frac{\partial}{\partial 
\theta} \sin\theta 
\frac{\partial}{\partial \theta} + \frac{1}{\sin^2\theta} \left( 
\frac{\partial^2}{\partial \psi^2} +   
\frac{\partial^2}{\partial \phi^2} - 2 \cos\theta   
\frac{\partial^2}{\partial \psi \partial \phi} \right)
\end{equation}
is the Laplace-Beltrami operator on the sphere.   
\par
Let us suppose that the Euler angles $\psi$, $\theta$ and $\phi$ are 
equal to $\psi'$,$\theta'$ and $\phi'$, respectively,  when the 
interaction-strength parameter $\eta$ is zero. Then the solution 
of the Fokker-Planck equation gives the conditional p.d.f. 
of the Euler angles 
\begin{eqnarray}
\label{fdef}
& & g(\psi,\theta,\phi;t|\psi',\theta',\phi') 
\nonumber \\   
& = & \sum_{j=0}^{\infty} \sum_{m=-j}^j \sum_{n=-j}^j 
\frac{2 j + 1}{32 \pi^2} D^j_{m,n}(\psi,\theta,\phi) 
\left\{ D^j_{m,n}(\psi',\theta' ,\phi') \right\}^* 
{\rm e}^{-j(j+1)\eta^2 D t}. \nonumber \\ 
\end{eqnarray}
Here $D^j_{m,n}$ is Wigner's D function\cite{LL}
\begin{equation}
D^j_{m,n}(\psi,\theta,\phi) = 
{\rm e}^{i m \phi} d^j_{m,n}(\theta) {\rm e}^{i n \psi},
\end{equation}
where
\begin{equation}
d^j_{m,n}(\theta) = \sqrt{\frac{(j + m)! (j - m)!}{(j + n)! (j - n)!}} 
\cos^{m+n}(\theta/2) \sin^{m-n}(\theta/2) P^{(m-n,m+n)}_{j-m}(\cos\theta)
\end{equation}
with the Jacobi polynomials $P^{(a,b)}_k(x)$. Note that $j$ is an 
integer or a half odd integer ($j = 0,1/2,1,3/2,\cdots$ and 
$m,n = -j,-j+1,\cdots,j$).
\par
Under the assumption described above, the factor 
$({\rm tr}\Delta_{\gamma}(\eta)) ({\rm tr}\Delta_{ \gamma'}(0))$ 
in (\ref{dsum}) with $\eta'= 0$ can be replaced by the 
average $\langle \langle \ ({\rm tr}\Delta_{\gamma}(\eta)) ({\rm tr}
\Delta_{\gamma'}(0)) \ \rangle \rangle$ over the Brownian motion. 
Thus we can write the form factor as
\begin{eqnarray}
\label{dsum2}
& & K(\tau;\eta,0) \nonumber \\ 
& \sim & \frac{1}{T_H^2} \left\langle \sum_{\gamma,\gamma'}
\langle \langle \ ({\rm tr}\Delta_{\gamma}(\eta)) ({\rm tr}\Delta_{
\gamma'}(0))^* \ \rangle \rangle  
A_{\gamma} A_{\gamma'}^* {\rm e}^{i (S_{\gamma} - S_{\gamma'})/
\hbar}
\delta\left( \tau - \frac{T_{\gamma} + T_{\gamma'}}{2 T_H} \right)
\right\rangle. \nonumber \\ 
\end{eqnarray}
We shall evaluate the $\tau$ expansion of this semiclassical 
form factor, focusing on the systems with spin $S = 1/2$.
\par
Let us  calculate the leading term in the $\tau$ expansion by using 
Berry's diagonal approximation\cite{BERRY}. In Berry's approximation, 
one first considers the contributions 
from the pairs of identical periodic orbits $(\gamma,\gamma)$. 
The spin evolution matrix along $\gamma$ with $S = 1/2$ is given by 
\begin{equation}
\Delta_{\gamma}(\eta) = {\rm exp}\left(\phi \frac{i}{2} \sigma_z \right) 
 {\rm exp}\left(\theta \frac{i}{2} \sigma_x \right) 
 {\rm exp}\left(\psi \frac{i}{2} \sigma_z \right), 
\end{equation}
where
\begin{equation}
\label{pauli}
\sigma_x = \left( \begin{array}{cc} 0 & 1 \\ 1 & 0 \end{array} \right), 
\ \ \ \sigma_z = 
\left( \begin{array}{cc} 1 & 0 \\ 0 & -1 \end{array} \right)
\end{equation}
are the Pauli matrices. It follows that
\begin{equation}
{\rm tr}\Delta_{\gamma}(\eta) = 2 \cos\frac{\theta}{2} \cos\left\{ 
\frac{1}{2} ( \psi + \phi ) \right\}.
\end{equation}  
The average of the factor $({\rm tr}\Delta_{\gamma}(\eta)) 
({\rm tr}\Delta_{\gamma}(0))$ over the Brownian motion 
can be written as 
\begin{eqnarray}
\label{ave}
& & \langle \langle \ ({\rm tr}\Delta_{\gamma}(\eta)) ({\rm tr}
\Delta_{\gamma}(0)) \ \rangle \rangle \nonumber \\ 
& = & \int {\rm d}\omega  {\rm d}\omega' 
({\rm tr}\Delta_{\gamma}(\eta)) ({\rm tr}\Delta_{\gamma}(0)) 
g(\psi,\theta,\phi;T|\psi',\theta',\phi')  
p_0(\psi',\theta',\phi'), \nonumber \\ 
\end{eqnarray}
where $p_0$ is the p.d.f. of the Euler angles at $\eta = 0$. 
The integrals are defined as 
\begin{eqnarray}
\int {\rm d}\omega & = & \int_0^{4 \pi} 
{\rm d}\psi \int_0^{\pi} {\rm d}\theta \int_0^{4 \pi} 
{\rm d}\phi \sin\theta, \nonumber \\  
\int {\rm d}\omega'& = & \int_0^{4 \pi} 
{\rm d}\psi' \int_0^{\pi} {\rm d}\theta' \int_0^{4 \pi} 
{\rm d}\phi' \sin\theta'
\end{eqnarray}
and $T = T_{\gamma}$ is the period of $\gamma$. 
\par
For the transition within the GSE universality 
class (the GSE to GSE transition), 
we employ the uniform "initial distribution"
\begin{equation}
\label{pzero}
p_0(\psi,\theta,\phi) = \frac{1}{32 \pi^2},
\end{equation}
since it yields the spectral form factor of the GSE 
class\cite{MHBHA05,BK,KEPPELER,BH}. The uniform distribution at 
$\eta=0$ implies that the spin is under the influence of additional 
interactions apart from the interaction described by (\ref{ham}). 
Putting (\ref{pzero}) into (\ref{ave}), we obtain 
\begin{eqnarray}
& & \langle \langle \ ({\rm tr}\Delta_{\gamma}(\eta)) ({\rm tr}\Delta_{
\gamma}(0)) \ \rangle \rangle  
\nonumber \\ 
& = & \frac{1}{32 \pi^2} \int {\rm d}\omega  {\rm d}\omega' 
({\rm tr}\Delta_{\gamma}(\eta)) ({\rm tr}\Delta_{ \gamma}(0)) 
g(\psi,\theta,\phi;T|\psi',\theta',\phi')  
\nonumber \\ 
& = & \frac{1}{32 \pi^2} \int {\rm d}\omega  {\rm d}\omega' 
\left\{ D^{1/2}_{1/2,1/2}(\psi,\theta,\phi) + 
D^{1/2}_{-1/2,-1/2}(\psi,\theta,\phi) \right\}^* \nonumber \\  
& & \times \left\{ D^{1/2}_{1/2,1/2}(\psi',\theta',\phi') + 
D^{1/2}_{-1/2,-1/2}(\psi',\theta',\phi') \right\}    
g(\psi,\theta,\phi;T|\psi',\theta',\phi').  
\nonumber \\ 
\end{eqnarray} 
Therefore, using the definition (\ref{fdef}) of $g$ and the 
orthogonality relation
\begin{equation}
\label{ort}
\int {\rm d}\omega \left\{ D^{j}_{m,n}(\psi,\theta,\phi) \right\}^*  
D^{j'}_{m',n'}(\psi,\theta,\phi) = \frac{32 \pi^2}{2 j + 1} 
\delta_{j,j'} \delta_{m,m'} \delta_{n,n'}, 
\end{equation}
we can readily find
\begin{equation}
\langle \langle \ ({\rm tr}\Delta_{\gamma}(\eta)) ({\rm tr}\Delta_{
\gamma}(0)) \ \rangle \rangle  
= {\rm e}^{-(3/4) a T} 
\end{equation}
with
\begin{equation}
a = \eta^2 D.
\end{equation}
Here the interaction-strength parameter $\eta$ is scaled 
so that $a T$ remains finite in the semiclassical 
limit $\hbar \rightarrow 0$. In order to take a step 
further, we need Hannay and Ozorio de Almeida (HOdA)'s sum 
rule\cite{HODA}
\begin{equation}
\label{hoda}
\frac{1}{T_H^2} \left\langle \sum_{\gamma}\left| A_{\gamma}\right|^2
\delta \left( \tau - \frac{T_{\gamma}}{T_{H}} \right) \right\rangle
= \tau,
\end{equation}
which results from the ergodicity of the system. Using this 
sum rule, we find the contribution to the form factor as
\begin{eqnarray}
K_{(\gamma,\gamma)}(\tau;\eta,0) & = &  
\frac{1}{T_H^2} \left\langle 
\sum_{\gamma}\left| A_{\gamma}\right|^2 \delta
\left( \tau - \frac{T_{\gamma}}{T_{H}}\right) \right\rangle 
\langle \langle 
\ ({\rm tr}\Delta_{\gamma}(\eta)) ({\rm tr}\Delta_{
\gamma}(0)) \ 
\rangle \rangle \nonumber \\ 
& = & \tau {\rm e}^{-(3/4)a T}.  
\end{eqnarray}
The second contribution to Berry's diagonal approximation 
comes from the pairs $(\gamma,{\bar \gamma})$, where a bar 
denotes time reversal. Noting 
\begin{eqnarray}
\Delta_{\bar \gamma}(\eta) & = & 
\{ \Delta_{\gamma}(\eta) \}^{-1}  
\nonumber \\ 
& = & 
\left( \begin{array}{cc} 0 & 1 \\ -1 & 0 \end{array} \right) 
\left\{ \Delta_{\gamma}(\eta) \right\}^{\rm T} 
\left( \begin{array}{cc} 0 & -1 \\ 1 & 0 \end{array} \right), 
\end{eqnarray}
where $\left\{ \Delta_{\gamma}(\eta) \right\}^{\rm T}$ is 
the transpose of $\Delta_{\gamma}(\eta)$, we find
\begin{equation}
\label{rev}
{\rm tr}\Delta_{\bar \gamma}(\eta) = 
{\rm tr}\Delta_{\gamma}(\eta).
\end{equation}
Therefore we can similarly obtain a contribution
\begin{eqnarray}
K_{(\gamma,{\bar \gamma})}(\tau;\eta,0) & = &  
\frac{1}{T_H^2} \left\langle 
\sum_{\gamma}\left| A_{\gamma}\right|^2 \delta
\left( \tau - \frac{T_{\gamma}}{T_{H}}\right) \right\rangle 
\langle \langle 
\ ({\rm tr}\Delta_{\gamma}(\eta)) ({\rm tr}\Delta_{\bar \gamma}
(0)) \ 
\rangle \rangle \nonumber \\ 
& = & \tau {\rm e}^{-(3/4)a T}.  
\end{eqnarray}
Thus the total sum of the contributions to the diagonal 
approximation is 
\begin{equation} 
\label{kdiag}
K_{\rm diag}(\tau) = K_{(\gamma,\gamma)}(\tau;\eta,0) +   
K_{(\gamma,{\bar \gamma})}(\tau;\eta,0) = 
2 \tau {\rm e}^{-(3/4)a T}.  
\end{equation} 

\section{Off-diagonal Contributions}
\setcounter{equation}{0}
\renewcommand{\theequation}{3.\arabic{equation}}

We are now in a position to calculate the off-diagonal
contributions, restricting ourselves to the systems with 
two degrees of freedom ($f = 2$). Encounters of periodic 
orbits play the major role in identifying the leading terms. An 
encounter is a set of orbit segments which 
come close to each other in the phase space. Long 
periodic orbits have encounters of the order 
of the Ehrenfest time $T_E$. In the semiclassical 
limit $\hbar \rightarrow 0$, $T_E$ logarithmically 
diverges. However, as the period $T$ is of the 
order of the Heisenberg time $T_H$, which more 
rapidly diverges, $T_E$ remains vanishingly small 
compared with the period. Therefore the periodic orbit 
mostly goes along loops in the phase space and occasionally visit 
encounters. As the leading terms are expected to result 
from the periodic orbit pairs $(\gamma,\gamma')$ which 
are close to each other or mutually almost time reversed, 
we can suppose that $\gamma'$ is almost identical to $\gamma$ 
or ${\bar \gamma}$ on the loops but differently connected in 
the encounters. 
\par
Let us consider such a periodic orbit pair $\alpha = 
(\gamma,\gamma')$ in the phase space. Within each 
encounter, a Poincar{\'e} section ${\cal P}$ orthogonal to 
the orbit $\gamma$ can be introduced. Suppose that 
$\gamma$ pierces ${\cal P}$ within the $r$-th encounter. 
If $l_r$ segments of $\gamma$ are contained in the $r$-th 
encounter, there are $l_r$ piercing points on ${\cal P}$. 
The displacement $\delta {\bf x}$ between such points can 
be spanned as $\delta {\bf x} = s {\hat e}_s 
+ u {\hat e}_u$. Here pairwise normalised vectors 
${\hat e}_s$ and ${\hat e}_u$ have directions along 
the stable and unstable manifolds, respectively.
Therefore, if one reference piercing point is chosen as 
the origin, each of other piercing points is 
identified by a coordinate pair $(s,u)$. As a result, if 
$\gamma$ has $L$ loops and $V$ encounters, $\sum_{r=1}^V 
(l_r - 1) = L - V$ coordinate pairs $(s_j,u_j)$, $j = 
1,2,\cdots,L-V$ are necessary to identify the piercing 
points of $\gamma$. 
\par
Let us denote by $T_j$ the duration on the $j$-th loop and by 
$t_r$ the duration of the $r$-th encounter. Then the total 
duration of the encounters is
\begin{equation}
t_{\alpha} \equiv \sum_{r=1}^V l_r t_r
\end{equation}
and the period is
\begin{equation} 
T = \sum_{j=1}^L T_j + t_{\alpha}. 
\end{equation}
Ergodicity can be employed to estimate the number of encounters 
as\cite{MHBHA04,MHBHA05,SM,SN}
\begin{equation}
\label{nume}
\int {\rm d}{\bf u}{\rm d}{\bf s}
\int_0^{T-t_{\alpha}} {\rm d}T_1
\int_0^{T-t_{\alpha}-T_1} {\rm d}T_2 \cdots
\int_0^{T-t_{\alpha}-T_1-T_2-\cdots-T_{L-2}} {\rm d}T_{L-1}
\ Q_{\alpha},
\end{equation}
where the integration measures are given by
\begin{equation}
{\rm d}{\bf u} = \prod_{j=1}^{L-V} {\rm d}u_j,
\ \ \ {\rm d}{\bf s} = \prod_{j=1}^{L-V} {\rm d}s_j
\end{equation}
and
\begin{equation}
Q_{\alpha} = \frac{T}{N_{\alpha} \
\prod_{r=1}^V t_r \ \Omega^{L-V}}.
\end{equation}
Here $N_{\alpha}$ is a combinatorial factor 
chosen such that overcountings are avoided. 
\par
Now the contribution to the form factor from the orbit pair 
$\alpha = (\gamma,\gamma')$ and its counterpart $(\gamma,{\bar \gamma'})$ 
can be readily derived. Referring to (\ref{dsum2}) and (\ref{nume}) 
and taking account of ({\ref{rev}), we find that 
such a contribution is   
\begin{eqnarray}
K_{\alpha}(\tau)
& = & 2 \tau \int {\rm d}{\bf u}{\rm d}{\bf s}
\int_0^{T-t_{\alpha}} {\rm d}T_1
\int_0^{T-t_{\alpha}-T_1} {\rm d}T_2 \cdots
\int_0^{T-t_{\alpha}-T_1-T_2-\cdots-T_{L-2}} {\rm d}T_{L-1}
\nonumber \\ 
& \times & Q_{\alpha} 
\langle \langle \ ({\rm tr}\Delta_{\gamma}(\eta)) ({\rm tr}\Delta_{
\gamma'}(0)) \ \rangle \rangle  
{\rm e}^{i \Delta S/\hbar}, 
\end{eqnarray}
where the action difference $\Delta S \equiv S_{\gamma} - 
S_{\gamma'}$ is given by $\Delta S = \sum_{j=1}^{L-V} 
u_j s_j$\cite{MHBHA04, MHBHA05,SM}. 
\par
In order to obtain a semiclassical result, we need to expand 
$K_{\alpha}(\tau)$ in $t_r$'s and extract the terms in which 
all $t_r$'s mutually cancel. Since extra factors $\hbar$ appear 
or rapid oscillations take place in the limit $\hbar \to 0$, 
the other terms should be neglected\cite{MHBHA04,MHBHA05,SM}. 
The off-diagonal contribution to the semiclassical form factor 
is thus derived as
\begin{equation}
\label{koff}
K_{\rm off}(\tau)
= \sum_{\alpha} 
\frac{2 \tau^2 T_H}{N_{\alpha}} \left( \frac{2}{T_H} \right)^{L-V}
\left. \frac{\partial^V}{\partial t_1 \partial t_2 \cdots \partial t_V} 
\Phi(t_1,t_2,\cdots,t_V) \right|_{t_1 = t_2 = \cdots = t_V = 0},
\end{equation}
where
\begin{eqnarray}
& & \Phi(t_1,t_2,\cdots,t_V) \nonumber \\ 
& = & \int_0^{T-t_{\alpha}} {\rm d}T_1
\int_0^{T - t_{\alpha}-T_1} {\rm d}T_2 \cdots
\int_0^{T - t_{\alpha} -T_1-T_2-\cdots-T_{L-2}} 
{\rm d}T_{L-1} \nonumber \\ 
& \times & 
\langle \langle \ ({\rm tr}\Delta_{\gamma}(\eta)) ({\rm tr}\Delta_{
\gamma'}(0)) \ \rangle \rangle. \nonumber \\  
\end{eqnarray}

\subsection{Sieber-Richter Term}

In this and next subsection we consider the $\tau$ expansion of 
the above formula (\ref{koff}). Mathematica was used to assist 
the computations. Each term of (\ref{koff}) is of 
order $\tau^n$ with $n = L-V+1$. Let us first consider the second order 
term ($n = 2$). The relevant pairs $\alpha = (\gamma,\gamma')$ 
has two loops ($L = 2$) and one encounter $(V = 1)$. Such periodic 
orbit pairs were identified by Sieber and Richter and thus called SR 
(Sieber-Richter) pairs\cite{SR}. An SR pair is schematically depicted 
in Figure 1. 

\begin{figure}[h!]
\epsfxsize=14cm
\centerline{\epsfbox{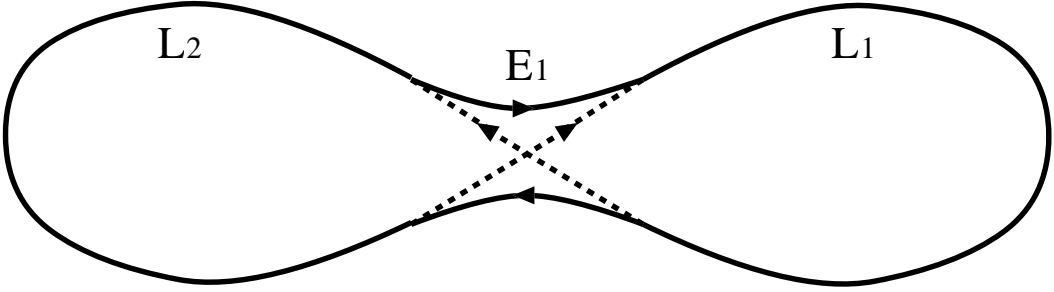}}
\caption{The periodic orbit pair contributing to the second order term}
\end{figure}

In Figure 1, $L_1$ and $L_2$ are loops and $E_1$ is an encounter. 
In the encounter, $\gamma$ and $\gamma'$ are depicted by solid curves 
and dashed lines, respectively, and each arrow shows the direction of the 
motion. We can symbolically write the periodic orbits as
\begin{equation} 
\gamma = {\bar E_1} L_2 E_1 L_1, \ \ \ \gamma' = {\bar E'_1} {\bar L'_2} 
E'_1 L'_1, 
\end{equation}
so that the spin evolution matrices are
\begin{equation} 
\Delta_\gamma 
= (\Delta_{E_1})^{-1} \Delta_{L_2} \Delta_{E_1} \Delta_{L_1}, 
\ \ \ \Delta_{\gamma'} = (\Delta_{E'_1})^{-1} 
(\Delta_{L'_2})^{-1} \Delta_{E'_1} 
\Delta_{L'_1}. 
\end{equation}
A spin evolution matrix $\Delta_{\chi}$ along a segment $\chi$ of a 
periodic orbit is given by (\ref{ev}) and can be expressed as 
\begin{equation}
\Delta_{\chi} = {\rm exp}\left(\phi_{\chi} \frac{i}{2} \sigma_z \right) 
 {\rm exp}\left(\theta_{\chi} \frac{i}{2} \sigma_x \right) 
 {\rm exp}\left(\psi_{\chi} \frac{i}{2} \sigma_z \right)
\end{equation}
in terms of a set of the Euler angles $\omega_{\chi} = 
(\psi_{\chi},\theta_{\chi},\phi_{\chi})$. The Pauli matrices 
$\sigma_x$ and $\sigma_z$ are defined in (\ref{pauli}). The 
corresponding integral over the Euler angles is defined as
\begin{equation}
\int {\rm d}\omega_{\chi} = \int_0^{4 \pi} {\rm d}\psi_{\chi} 
\int_0^{\pi} {\rm d}\theta_{\chi} \int_0^{4 \pi} 
{\rm d}\phi_{\chi} \sin\theta_{\chi}.     
\end{equation}
Moreover we denote the durations of $L_1$, $L_2$ and $E_1$ by 
$T_1$, $T_2$ and $t_1$, respectively. Using the above notations, 
we evaluate the average of $({\rm tr}\Delta_{\gamma}(\eta)) 
({\rm tr}\Delta_{\gamma'}(0))$ as 
\begin{eqnarray} 
& & \langle \langle \ ({\rm tr}\Delta_{\gamma}(\eta)) ({\rm tr}\Delta_{
\gamma'}(0)) \ \rangle \rangle \nonumber \\ 
& = & 
\frac{1}{(32 \pi^2)^3} \int {\rm d}\omega_{L_1} 
{\rm d}\omega_{L_2} {\rm d}\omega_{E_1} 
\int {\rm d}\omega_{L'_1} {\rm d}\omega_{L'_2} {\rm d}\omega_{E'_1} 
\nonumber \\ & \times & 
{\rm tr}((\Delta_{E_1})^{-1} \Delta_{L_2} \Delta_{E_1} \Delta_{L_1}) 
{\rm tr}((\Delta_{E'_1})^{-1} (\Delta_{L'_2})^{-1} \Delta_{E'_1} \Delta_{L'_1})
\nonumber \\ & \times & g(\omega_{L_1};T_1|\omega_{L'_1})    
g(\omega_{L_2};T - T_1 - 2 t_1|\omega_{L'_2})    
g(\omega_{E_1};t_1|\omega_{E'_1})  
\nonumber \\ 
& =  & 
\frac{1}{4} {\rm e}^{-(3/4) a T} \left( 
{\rm e}^{(3/2) a t_1} - 3 {\rm e}^{-(1/2) a t_1} 
\right),
\end{eqnarray}
so that
\begin{equation}
\Phi(t_1) = \frac{T - 2 t_1}{4} {\rm e}^{-(3/4) a T} 
\left( {\rm e}^{(3/2) a t_1} 
- 3 {\rm e}^{-(1/2) a t_1} 
\right).
\end{equation}
Due to the equivalence of the segments $E_1$ and ${\bar E_1}$, 
we need to choose the combinatorial factor $N_{\alpha}$ as 
$N_{\rm SR} = 2$\cite{HMBH04}. Consequently we find the contribution 
from the SR pairs to the form factor as 
\begin{equation}
\label{ksr}
K_{\rm SR}(\tau) = \frac{4 \tau^2}{N_{\rm SR}} \left. 
\frac{\partial}{\partial t_1} 
\Phi(t_1) \right|_{t_1 = 0} = 2 \tau^2 {\rm e}^{-(3/4) a T} 
\left( 1 + \frac{3}{4} a T \right). 
\end{equation}

\subsection{Third Order Term}

Next we consider the third order term ($n = L - V + 1 = 3$). 
It is known that the periodic orbit pairs contributing to 
the third order term are classified into five types: 
aas, api, ppi, ac and pc\cite{HMBH04}. These five types 
are depicted in Figure 2. 

\begin{figure}[h!]
\epsfxsize=14cm
\centerline{\epsfbox{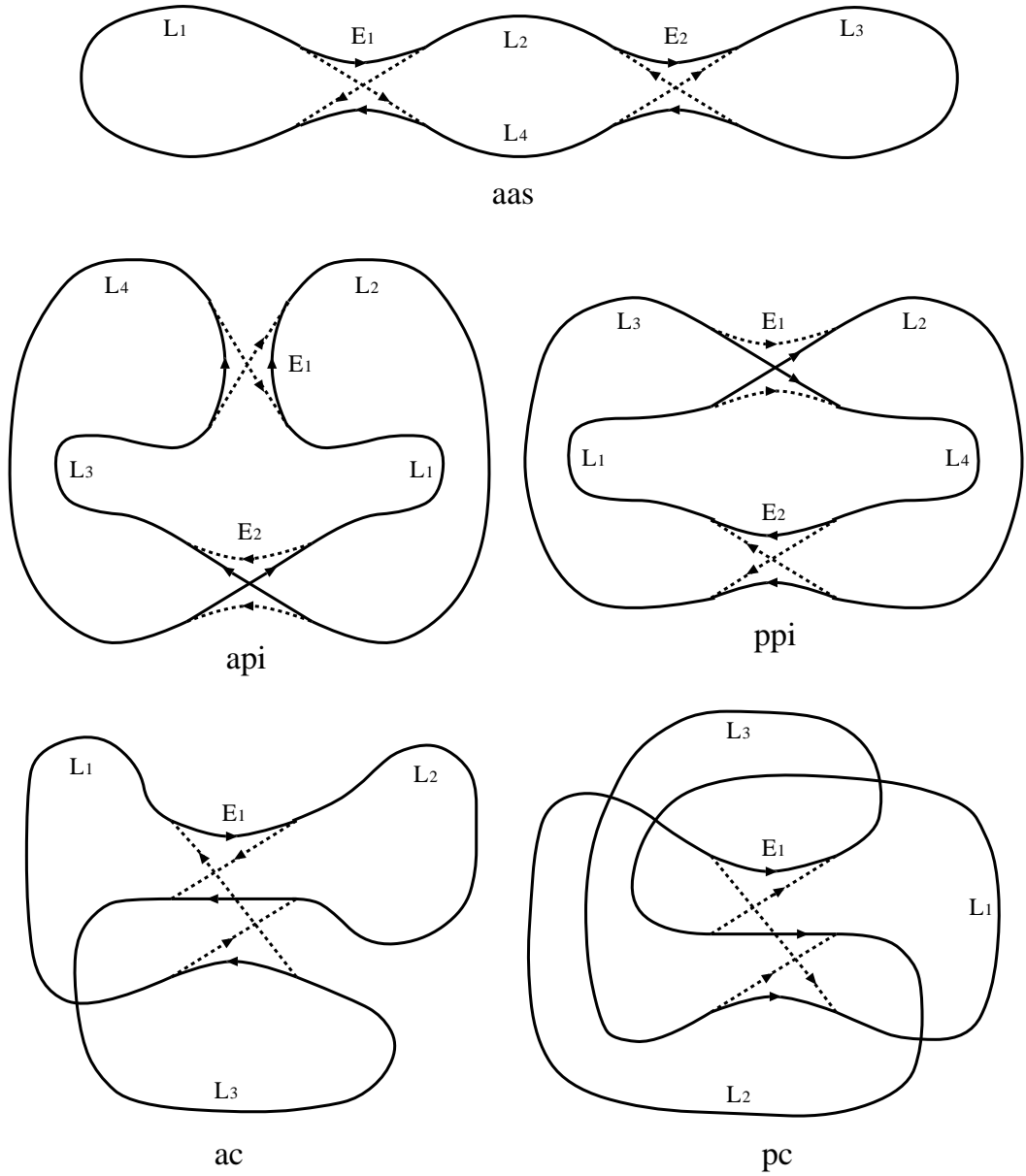}}
\caption{The periodic orbit pairs contributing to the third order term}
\end{figure}

As is seen from Figure 2, each of aas, api and ppi orbit pairs has 
four loops ($L = 4$) and two encounters ($V = 2$). The durations of the 
loops $L_j$ ($j = 1,2,3,4$) and the encounters $E_l$ ($l = 1,2$) are 
denoted by $T_j$ and $t_l$, respectively. The combinatorial factors 
$N_{\alpha}$ are known to be given by $N_{\rm aas} = 2$, $N_{\rm api} 
= 2$ and $N_{\rm ppi} = 4$\cite{HMBH04}.          
\par
On the other hand, each of ac and pc orbit pairs has three loops 
($L = 3$) and one encounter ($V = 1$). The times elapsed on the 
loops $L_j$ ($j = 1,2,3$) and on the encounter $E_1$ are denoted 
by $T_j$ and $t_1$, respectively. The combinatorial factors 
$N_{\alpha}$ are $N_{\rm ac} = 1$ and $N_{\rm pc} = 3$\cite{HMBH04}.   
\par
In the following we calculate the contribution to the 
form factor $K_{\alpha}(\tau)$ from each of the five types: 
$\alpha = {\rm aas}, {\rm api}, {\rm ppi}, {\rm ac}$ 
and ${\rm pc}$.  
\par
\medskip
\noindent
(1) aas orbit pairs ($N_{\rm aas} = 2$) 
\par
\medskip
\noindent
\begin{eqnarray} 
& & \langle \langle \ ({\rm tr}\Delta_{\gamma}(\eta)) ({\rm tr}\Delta_{
\gamma'}(0)) \ \rangle \rangle \nonumber \\ 
& = & 
\frac{1}{(32 \pi^2)^6} 
\int {\rm d}\omega_{L_1} {\rm d}\omega_{L_2} 
{\rm d}\omega_{L_3} {\rm d}\omega_{L_4} 
{\rm d}\omega_{E_1} {\rm d}\omega_{E_2} 
\nonumber \\ & \times & 
\int {\rm d}\omega_{L'_1} {\rm d}\omega_{L'_2} 
{\rm d}\omega_{L'_3} {\rm d}\omega_{L'_4} 
{\rm d}\omega_{E'_1} {\rm d}\omega_{E'_2} 
\nonumber \\ & \times & 
{\rm tr}(\Delta_{E_1} \Delta_{L_2} \Delta_{E_2} \Delta_{L_3} (\Delta_{E_2})^{-1} 
\Delta_{L_4} (\Delta_{E_1})^{-1} \Delta_{L_1}) 
\nonumber \\ & \times & 
{\rm tr}(\Delta_{E'_1} (\Delta_{L'_4})^{-1} \Delta_{E'_2} \Delta_{L'_3} (\Delta_{E'_2})^{-1} 
(\Delta_{L'_2})^{-1} (\Delta_{E'_1})^{-1} \Delta_{L'_1})
\nonumber \\ & \times & 
g(\omega_{L_1};T_1|\omega_{L'_1})    
g(\omega_{L_2};T_2|\omega_{L'_2})    
g(\omega_{L_3};T_3|\omega_{L'_3})    
\nonumber \\ & \times & 
g(\omega_{L_4};T - 2 t_1 - 2 t_2 - T_1 - T_2 - T_3|\omega_{L'_4})    
\nonumber \\ & \times & 
g(\omega_{E_1};t_1|\omega_{E'_1})  
g(\omega_{E_2};t_2|\omega_{E'_2}) \nonumber \\ 
& = & \frac{1}{16} {\rm e}^{-(3/4) a T} 
\left( 
{\rm e}^{(3/2) a t_1} - 3 {\rm e}^{-(1/2) a t_1} \right) 
\left( 
{\rm e}^{(3/2) a t_2} - 3 {\rm e}^{-(1/2) a t_2} \right). 
\end{eqnarray}
Therefore
\begin{eqnarray}
& & \Phi(t_1,t_2) \nonumber \\ 
& = & \frac{(T - 2 t_1 - 2 t_2)^3}{96} {\rm e}^{-(3/4) a T} 
\left( 
{\rm e}^{(3/2) a t_1} - 3 {\rm e}^{-(1/2) a t_1} \right) 
\left( 
{\rm e}^{(3/2) a t_2} - 3 {\rm e}^{-(1/2) a t_2} \right), 
\nonumber \\ 
\end{eqnarray}
so that
\begin{eqnarray}
K_{\rm aas}(\tau) & = & \frac{8 \tau^2}{N_{\rm aas} T_H} \left. 
\frac{\partial^2}{\partial t_1 \partial t_2} 
\Phi(t_1,t_2) \right|_{t_1 = t_2 = 0} \nonumber \\ 
& = & 4 \tau^3 {\rm e}^{-(3/4) a T} 
\left\{ 1 + \frac{3}{4} a T  + \frac{3}{32} (a T)^2 \right\}. 
\end{eqnarray}
\par
\medskip
\noindent
(2) api orbit pairs ($N_{\rm api} = 2$) 
\par
\medskip
\noindent
\begin{eqnarray} 
& & \langle \langle \ ({\rm tr}\Delta_{\gamma}(\eta)) ({\rm tr}\Delta_{
\gamma'}(0)) \ \rangle \rangle \nonumber \\ 
& = & 
\frac{1}{(32 \pi^2)^6} 
\int {\rm d}\omega_{L_1} {\rm d}\omega_{L_2} 
{\rm d}\omega_{L_3} {\rm d}\omega_{L_4} 
{\rm d}\omega_{E_1} {\rm d}\omega_{E_2} 
\nonumber \\ & \times & 
\int {\rm d}\omega_{L'_1} {\rm d}\omega_{L'_2} 
{\rm d}\omega_{L'_3} {\rm d}\omega_{L'_4} 
{\rm d}\omega_{E'_1} {\rm d}\omega_{E'_2} 
\nonumber \\ & \times & 
{\rm tr}(\Delta_{E_1} \Delta_{L_2} \Delta_{E_2} \Delta_{L_3} \Delta_{E_1} 
\Delta_{L_4} (\Delta_{E_2})^{-1} \Delta_{L_1}) 
\nonumber \\ & \times & 
{\rm tr}(\Delta_{E'_1} \Delta_{L'_2} \Delta_{E'_2} (\Delta_{L'_4})^{-1} 
(\Delta_{E'_1})^{-1} (\Delta_{L'_1})^{-1} \Delta_{E'_2} \Delta_{L'_3})
\nonumber \\ & \times & 
g(\omega_{L_1};T_1|\omega_{L'_1})    
g(\omega_{L_2};T_2|\omega_{L'_2})    
g(\omega_{L_3};T_3|\omega_{L'_3})    
\nonumber \\ & \times & 
g(\omega_{L_4};T - 2 t_1 - 2 t_2 - T_1 - T_2 - T_3|\omega_{L'_4})    
\nonumber \\ & \times & 
g(\omega_{E_1};t_1|\omega_{E'_1})  
g(\omega_{E_2};t_2|\omega_{E'_2}) \nonumber \\   
& = & \frac{1}{16} {\rm e}^{-(3/4) a T} \left( 
{\rm e}^{(3/2) a t_1} - 3 {\rm e}^{-(1/2) a t_1} \right) 
\left( 
{\rm e}^{(3/2) a t_2} - 3 {\rm e}^{-(1/2) a t_2} \right). 
\end{eqnarray}
It follows that
\begin{equation}
K_{\rm api}(\tau) = 4 \tau^3 {\rm e}^{-(3/4) a T} 
\left\{ 1 + \frac{3}{4} a T  + \frac{3}{32} (a T)^2 \right\}. 
\end{equation}
\par
\medskip
\noindent
(3) ppi orbit pairs ($N_{\rm ppi} = 4$) 
\par
\medskip
\noindent
\begin{eqnarray} 
& & \langle \langle \ ({\rm tr}\Delta_{\gamma}(\eta)) ({\rm tr}\Delta_{
\gamma'}(0)) \ \rangle \rangle \nonumber \\ 
& = & 
\frac{1}{(32 \pi^2)^6} 
\int {\rm d}\omega_{L_1} {\rm d}\omega_{L_2} 
{\rm d}\omega_{L_3} {\rm d}\omega_{L_4} 
{\rm d}\omega_{E_1} {\rm d}\omega_{E_2} 
\nonumber \\ & \times & 
\int {\rm d}\omega_{L'_1} {\rm d}\omega_{L'_2} 
{\rm d}\omega_{L'_3} {\rm d}\omega_{L'_4} 
{\rm d}\omega_{E'_1} {\rm d}\omega_{E'_2} 
\nonumber \\ & \times & 
{\rm tr}(\Delta_{E_1} \Delta_{L_2} \Delta_{E_2} \Delta_{L_3} \Delta_{E_1} 
\Delta_{L_4} \Delta_{E_2} \Delta_{L_1}) 
\nonumber \\ & \times & 
{\rm tr}(\Delta_{E'_1} \Delta_{L'_2} \Delta_{E'_2} \Delta_{L'_1} 
\Delta_{E'_1} \Delta_{L'_4} \Delta_{E'_2} \Delta_{L'_3})
\nonumber \\ & \times & 
g(\omega_{L_1};T_1|\omega_{L'_1})    
g(\omega_{L_2};T_2|\omega_{L'_2})    
g(\omega_{L_3};T_3|\omega_{L'_3})    
\nonumber \\ & \times & 
g(\omega_{L_4};T - 2 t_1 - 2 t_2 - T_1 - T_2 - T_3|\omega_{L'_4})    
\nonumber \\ & \times & 
g(\omega_{E_1};t_1|\omega_{E'_1})  
g(\omega_{E_2};t_2|\omega_{E'_2}) \nonumber \\  
& = & \frac{1}{16} {\rm e}^{-(3/4) a T} \left( 
{\rm e}^{(3/2) a t_1} - 3 {\rm e}^{-(1/2) a t_1} \right) 
\left( 
{\rm e}^{(3/2) a t_2} - 3 {\rm e}^{-(1/2) a t_2} \right). 
\end{eqnarray}
It follows that
\begin{equation}
K_{\rm ppi}(\tau) = 2 \tau^3 {\rm e}^{-(3/4) a T} 
\left\{ 1 + \frac{3}{4} a T  + \frac{3}{32} (a T)^2 \right\}. 
\end{equation}
\par
\medskip
\noindent
(4) ac orbit pairs ($N_{\rm ac} = 1$) 
\par
\medskip
\noindent
\begin{eqnarray} 
& & \langle \langle \ ({\rm tr}\Delta_{\gamma}(\eta)) ({\rm tr}\Delta_{
\gamma'}(0)) \ \rangle \rangle \nonumber \\ 
& = & 
\frac{1}{(32 \pi^2)^4} 
\int {\rm d}\omega_{L_1} {\rm d}\omega_{L_2} {\rm d}\omega_{L_3}  
{\rm d}\omega_{E_1} 
\int {\rm d}\omega_{L'_1} {\rm d}\omega_{L'_2} {\rm d}\omega_{L'_3}  
{\rm d}\omega_{E'_1} 
\nonumber \\ & \times & 
{\rm tr}(\Delta_{E_1} \Delta_{L_1} (\Delta_{E_1})^{-1} 
\Delta_{L_2} \Delta_{E_1} \Delta_{L_3}) 
{\rm tr}(\Delta_{E'_1} (\Delta_{L'_1})^{-1} (\Delta_{E'_1})^{-1} (\Delta_{L'_2})^{-1} 
\Delta_{E'_1} \Delta_{L'_3})
\nonumber \\ & \times & 
g(\omega_{L_1};T_1|\omega_{L'_1})    
g(\omega_{L_2};T_2|\omega_{L'_2})    
g(\omega_{L_3};T - 3 t_1 - T_1 - T_2|\omega_{L'_3})    
\nonumber \\ & \times & 
g(\omega_{E_1};t_1|\omega_{E'_1}) \nonumber \\  
& = & \frac{1}{4} {\rm e}^{-(3/4) a T} \left( 
2 {\rm e}^{-(3/2) a t_1} - {\rm e}^{(3/2) a t_1} \right). 
\end{eqnarray}
Therefore
\begin{equation}
\Phi(t_1) = \frac{(T - 3 t_1)^2}{8} {\rm e}^{-(3/4) a T} 
\left( 2 {\rm e}^{-(3/2) a t_1} - {\rm e}^{(3/2) a t_1} \right), 
\end{equation}
so that
\begin{equation}
K_{\rm ac}(\tau) = \frac{8 \tau^2}{N_{\rm ac} T_H} \left. 
\frac{\partial}{\partial t_1} \Phi(t_1) \right|_{t_1 = 0} 
= - 6  \tau^3 {\rm e}^{-(3/4) a T} 
\left( 1 + \frac{3}{4} a T \right). 
\end{equation}
\par
\medskip
\noindent
(5) pc orbit pairs ($N_{\rm pc} = 3$)
\par
\medskip
\noindent
\begin{eqnarray} 
& & \langle \langle \ ({\rm tr}\Delta_{\gamma}(\eta)) ({\rm tr}\Delta_{
\gamma'}(0)) \ \rangle \rangle \nonumber \\ 
& = & 
\frac{1}{(32 \pi^2)^4} 
\int {\rm d}\omega_{L_1} {\rm d}\omega_{L_2} {\rm d}\omega_{L_3}  
{\rm d}\omega_{E_1} 
\int {\rm d}\omega_{L'_1} {\rm d}\omega_{L'_2} {\rm d}\omega_{L'_3}  
{\rm d}\omega_{E'_1} 
\nonumber \\ & \times & 
{\rm tr}(\Delta_{E_1} \Delta_{L_1} \Delta_{E_1} 
\Delta_{L_2} \Delta_{E_1} \Delta_{L_3}) 
{\rm tr}(\Delta_{E'_1} \Delta_{L'_1} \Delta_{E'_1} 
\Delta_{L'_3} \Delta_{E'_1} \Delta_{L'_2})
\nonumber \\ & \times & 
g(\omega_{L_1};T_1|\omega_{L'_1})    
g(\omega_{L_2};T_2|\omega_{L'_2})    
g(\omega_{L_3};T - 3 t_1 - T_1 - T_2|\omega_{L'_3})    
\nonumber \\ & \times & 
g(\omega_{E_1};t_1|\omega_{E'_1}) \nonumber \\  
& = & \frac{1}{4} {\rm e}^{-(3/4) a T} \left( 
2 {\rm e}^{-(3/2) a t_1} - {\rm e}^{(3/2) a t_1} \right). 
\end{eqnarray}
It follows that
\begin{equation}
K_{\rm pc}(\tau) = - 2  \tau^3 {\rm e}^{-(3/4) a T} 
\left( 1 + \frac{3}{4} a T \right). 
\end{equation}
\par
\medskip
\noindent
Putting the above results together, we obtain the third order 
contribution to the form factor
\begin{eqnarray}
\label{k3rd}
K_{\rm 3rd}(\tau) & = & K_{\rm aas}(\tau) +  K_{\rm api}(\tau) 
+  K_{\rm ppi}(\tau) +  K_{\rm ac}(\tau) +  K_{\rm pc}(\tau) 
\nonumber \\ 
& = & 2 \tau^3 {\rm e}^{-(3/4) a T} \left\{ 1 + \frac{3}{4} 
a T + \frac{15}{32} (a T)^2 \right\}.
\end{eqnarray} 
Hence the semiclassical form factor up to the third order 
is calculated from (\ref{kdiag}), ({\ref{ksr}) and 
(\ref{k3rd}) as 
\begin{eqnarray}
\label{ksc}
K_{\rm SC}(\tau) & = & K_{\rm diag}(\tau) +  K_{\rm SR}(\tau) 
+  K_{\rm 3rd}(\tau)  
\nonumber \\ 
& = &  2 \tau {\rm e}^{-(3/4) a T} \left[  
1 + \left( 1 + \frac{3}{4} a T \right) \tau +  
 \left\{ 1 + \frac{3}{4} a T + \frac{15}{32} (a T)^2 \right\} 
\tau^2 \right]. \nonumber \\ 
\end{eqnarray} 

\section{Parametric Random Matrix Theory}
\setcounter{equation}{0}
\renewcommand{\theequation}{4.\arabic{equation}}

Parametric random matrix theory was originally invented 
by Dyson\cite{DYSON}. The quantum Hamiltonian of a 
time reversal invariant system with spin $1/2$ is 
simulated by an $N \times N$ self-dual real quaternion 
random matrix $H$. It is assumed to be a sum of a self-dual 
real quaternion matrix $H_0$ and a Gaussian random 
perturbation: the p.d.f. of $H$ is given by 
\begin{equation}
\label{prm}
P(H;\sigma | H_0) \ {\rm d}H \propto
{\rm exp}\left[- 2 \frac{{\rm Tr}\left\{ (H - {\rm e}^{-\sigma} H_0)^2
\right\}}{1 - {\rm e}^{- 2 \sigma}} \right] {\rm d}H
\end{equation}
with
\begin{equation}
{\rm d}H = \prod_{j=1}^N {\rm d}H_{jj} \prod_{j<l}^N \prod_{k=0}^3 
{\rm d}H^{(k)}_{jl}.
\end{equation}
Here $H^{(k)}_{jl}$ is the $k$-th component of the real quaternion 
$H_{jl}$. We are interested in the parametric motion of the matrix 
$H$ depending on the fictitious time parameter $\sigma$. 
\par
Let us write the eigenvalues of the self-dual real quaternion matrices 
$H$ and $H_0$ as $x_1,x_2,\cdots,x_N$ and $y_1,y_2,\cdots,
y_N$, respectively. Dyson derived the Fokker-Planck equation
\begin{equation}
\label{fpe}
\frac{\partial p}{\partial \sigma} = \sum_{j=1}^N 
\frac{\partial}{\partial x_j} \left(  
\frac{\partial W}{\partial x_j} p + \frac{1}{4} 
\frac{\partial p}{\partial x_j} \right) 
\end{equation}
with
\begin{equation}
W = \frac{1}{2} \sum_{j=1}^N (x_j)^2 - \sum_{j<l}^N \log|x_j - x_l|
\end{equation}
for the p.d.f. $p$ of the eigenvalues of $H$. 
\par
We denote by 
\begin{equation}
G(x_1,x_2,\cdots,x_N;\sigma |y_1,y_2,\cdots,y_N)
\end{equation}
the Green function solution of the Fokker-Planck equation (\ref{fpe}). 
Namely, $G$ with the measure $\prod_{j=1}^N {\rm d}x_j$ gives the p.d.f. 
of the eigenvalues of $H$ at $\sigma$ under the condition that 
$x_j = y_j$ ($j=1,2,\cdots,N$) at $\sigma = 0$. The limit $\sigma 
\to \infty$ of the Green function is given by the p.d.f. of the 
GSE eigenvalues
\begin{equation}
G(x_1,x_2,\cdots,x_N;\infty |y_1,y_2,\cdots,y_N)
=  p_{\rm GSE}(x_1,x_2,\cdots,x_N),
\end{equation}
where
\begin{equation}
\label{gse}
p_{\rm GSE}(x_1,x_2,\cdots,x_N) \propto {\rm e}^{- 4 W}. 
\end{equation}
\par
Let us choose the initial matrix $H_0$ as a GSE random matrix. 
Then the transition within the GSE symmetry class (the GSE to GSE 
transition) is realized. We define the dynamical (density-density) 
correlation function describing the correlation between the 
eigenvalues of $H$ and $H_0$ as
\begin{equation}
\kappa(x;\sigma|y) =  N^2 \frac{I(x;\sigma|y)}{I_0},
\end{equation}
where
\begin{eqnarray}
I(x_1;\sigma | y_1) & = &
\int_{-\infty}^{\infty} {\rm d}x_2
\int_{-\infty}^{\infty} {\rm d}x_3 \cdots
\int_{-\infty}^{\infty} {\rm d}x_N
\int_{-\infty}^{\infty} {\rm d}y_2
\int_{-\infty}^{\infty} {\rm d}y_3 \cdots
\int_{-\infty}^{\infty} {\rm d}y_N
\nonumber \\
& \times &
G( x_1,x_2,\cdots,x_N;\sigma | y_1,y_2,\cdots,y_N)
p_{\rm GSE}(y_1,y_2,\cdots,y_N)
\nonumber \\
\end{eqnarray}
and
\begin{equation}
I_0 = \int_{-\infty}^{\infty} {\rm d}x
\int_{-\infty}^{\infty} {\rm d}y
I(x;\sigma|y).
\end{equation}
\par
The asymptotic limit $N \to \infty$ of the dynamical
correlation function was evaluated by the method of 
supersymmetry\cite{SLA}. It can also be derived by using 
the properties of the Jack symmetric polynomials\cite{HA}. 
Let us note that the asymptotic eigenvalue density at $\sqrt{2 N} z$ 
($-1< z < 1$) is given by $\rho = \sqrt{2 N (1 - z^2)}/\pi$.
In terms of the new scaled variables $c,X$ and $Y$ defined as
\begin{equation}
\sigma = c/(\pi^2 \rho^2),  \ \ \
x  = \sqrt{2 N} z + (X/\rho), \ \ \  y = \sqrt{2 N} z + (Y/\rho),
\end{equation}
one obtains the asymptotic limit
\begin{eqnarray}
& & \frac{\kappa(x;\sigma|y)}{\rho^2} - 1 \sim {\bar \rho}(\xi;c) 
\nonumber \\ & \equiv & 
\frac{1}{2} \int_1^{\infty} {\rm d}u \int_{-1}^1 {\rm d}v_1 
\int_{-1}^1 {\rm d}v_2 \frac{(u^2-1) (u - v_1 v_2)^2}{\{2 u v_1 v_2 
- u^2 - (v_1)^2 - (v_2)^2 + 1\}^2} \nonumber \\ 
& \times & {\rm e}^{- c \{u^2 + (v_1)^2 + (v_2)^2 
- 2 (v_1)^2 (v_2)^2 - 1\}} \cos\{ 2 \pi \xi ( u - v_1 v_2)\}
\end{eqnarray}
with $\xi = X - Y$. The Fourier transform of the asymptotic limit  
\begin{equation}
K_{\rm RM}(\tau) =
\int_{-\infty}^{\infty} {\rm d}\xi \ {\rm e}^{i 2 \pi \tau \xi} \ {\bar
\rho}(\xi;c)
\end{equation}
gives the definition of the form factor. It can be written as
\begin{eqnarray}
K_{\rm RM}(\tau) & = &  
\frac{\tau^2}{2} \int_{1 - \tau}^1 {\rm d}v_1 
\int_{(1 - \tau)/v_1}^1 {\rm d}v_2 
\nonumber \\ & \times & \frac{(v_1 v_2 + \tau)^2 - 1}{ 
\{2 (v_1 v_2 + \tau) v_1 v_2 - (v_1 v_2 + \tau)^2 - (v_1)^2 - (v_2)^2 + 1\}^2} 
\nonumber \\     
& \times & {\rm e}^{-c \{(v_1 v_2 + \tau)^2 + (v_1)^2 + (v_2)^2 
- 2 (v_1)^2 (v_2)^2 - 1\}}
\end{eqnarray}
for $0 \leq \tau \leq 1$. In order to derive the $\tau$ 
expansion of $K_{\rm RM}(\tau)$, we introduce new integration 
variables $s_1$ and $s_2$ by 
\begin{equation}
\lambda_1 = 1 - \frac{\tau}{2} s_1, \ \ \ 
\lambda_1 \lambda_2 = 1 - \frac{\tau}{2} s_2.
\end{equation}    
Then we find
\begin{eqnarray}
& & K_{\rm RM}(\tau) = \frac{\tau}{8} \int_0^2 {\rm d}s_1 \int_{s_1}^2 
{\rm d}s_2 \ {\rm exp}\left\{  
- 2 \lambda \left( 1 - \frac{\tau}{2} s_2 \right) 
- \lambda \tau  \right\} \nonumber \\ & \times & 
{\rm exp}\left\{ \lambda \tau 
\frac{\displaystyle
\left( s_1 - \frac{\tau}{4} (s_1)^2 \right) 
\left( - s_1 + s_2 + \frac{\tau}{4} (s_1)^2 - 
\frac{\tau}{4} (s_2)^2 \right)} 
{\displaystyle 
\left( 1  - \frac{\tau}{2} s_1 \right)^2} 
\right\}
\nonumber \\ & \times & 
\frac{\displaystyle 
\left(1 - \frac{\tau}{2} s_1 \right)^3 
\left(2 - \frac{\tau}{2} s_2 + \tau \right) 
\left(1 - \frac{s_2}{2} \right)}
{\displaystyle \left\{
\left(s_1 - \frac{\tau}{4} (s_1)^2 \right)
\left( -s_1 + s_2 + \frac{\tau}{4} (s_1)^2 
- \frac{\tau}{4} (s_2)^2 \right) - 
\left(1 - \frac{\tau}{2} s_1 \right)^2 \right\}^2}, 
\nonumber \\ 
\end{eqnarray}
where $\lambda = c \tau$. Thus we can readily calculate 
the $\tau$ expansion (with fixed $\lambda$) from 
the Taylor expansion of the integrand as
\begin{equation}
K_{\rm RM}(\tau) = \frac{\tau}{8} {\rm e}^{- 2 \lambda} 
\left\{ 4 + (2 + 4 \lambda) \tau + \left( 
1 + 2 \lambda + \frac{10}{3} \lambda^2 \right) \tau^2 + \cdots \right\}.
\end{equation}
\par 
In order to compare this result with the semiclassical formula, 
we need to take account of the Kramers degeneracy, 
which means that all the eigenvalues have multiplicity two 
due to time reversal symmetry. Inclusion of the degeneracy 
yields a modified form factor  
\begin{eqnarray}
{\tilde K}_{\rm RM}(\tau) & = & 2 K_{\rm RM}(2 \tau) 
\nonumber \\ & = & 
2 \tau {\rm e}^{- 2 \lambda} 
\left\{1 + (1 + 2 \lambda) \tau + \left( 
1 + 2 \lambda + \frac{10}{3} \lambda^2 \right) \tau^2 + \cdots \right\}. 
\nonumber \\ 
\end{eqnarray} 
This is in agreement with the semiclassical formula (\ref{ksc}) up to 
the third order with an identification $\lambda = (3/8) a T$.   

\section{The GOE to GSE Transition}
\setcounter{equation}{0}
\renewcommand{\theequation}{5.\arabic{equation}}

If the spin evolution operator is represented by 
an identity matrix, the system is effectively 
spinless and the resulting spectral correlation 
belongs to the GOE universality class. Therefore, 
the crossover from the GOE class to the GSE class 
can be treated by introducing 
\begin{equation}
p_0(\psi,\theta,\phi) = \delta(\psi) 
\delta(\cos\theta - 1) \delta(\phi)
\end{equation}
as the "initial distribution" instead of (\ref{pzero}). 
In this section we investigate the GOE to GSE transition, 
focusing on the form factor $K(\tau, \eta,\eta)$, where $\eta'$ 
is equated with $\eta$. 
\par
As before, due to the relation (\ref{rev}), the 
contributions from the pairs $(\gamma,\gamma')$ 
and $(\gamma,{\bar \gamma}')$ are equal. Therefore, 
in order to calculate the form factor in the diagonal 
approximation, it suffices to treat the pairs 
$(\gamma,\gamma)$. The average over the 
Brownian motion can be evaluated as 
\begin{eqnarray}
& & \langle \langle \ ({\rm tr}\Delta_{\gamma}(\eta))^2
 \ \rangle \rangle \nonumber \\ 
& = & \int {\rm d}\omega  {\rm d}\omega' 
({\rm tr}\Delta_{\gamma}(\eta))^2 
g(\psi,\theta,\phi;T|\psi',\theta',\phi')  
p_0(\psi',\theta',\phi'), \nonumber \\ 
& = & \int {\rm d}\omega  
({\rm tr}\Delta_{\gamma}(\eta))^2 
g(\psi,\theta,\phi;T|0,0,0).  
\end{eqnarray}
Noting
\begin{equation}
({\rm tr}\Delta_{\gamma}(\eta))^2  = 
D^0_{0,0}(\psi,\theta,\phi)    
+ D^1_{-1,-1}(\psi,\theta,\phi)    
+ D^1_{0,0}(\psi,\theta,\phi)    
+ D^1_{1,1}(\psi,\theta,\phi)
\end{equation}    
and the orthogonality relation (\ref{ort}), we can readily find
\begin{equation}
\langle \langle \ ({\rm tr}\Delta_{\gamma}(\eta))^2  \ \rangle \rangle  
= 1 + 3 {\rm e}^{- 2 a T}. 
\end{equation}
Then, using the HOdA sum rule (\ref{hoda}), we find the contribution 
to the form factor
\begin{eqnarray}
K_{(\gamma,\gamma)}(\tau;\eta,\eta) & = &  
\frac{1}{T_H^2} \left\langle 
\sum_{\gamma}\left| A_{\gamma}\right|^2 \delta
\left( \tau - \frac{T_{\gamma}}{T_{H}}\right) \right\rangle 
\langle \langle 
\ ({\rm tr}\Delta_{\gamma}(\eta))^2 \ 
\rangle \rangle \nonumber \\ 
& = & \tau (1 + 3 {\rm e}^{- 2 a T}),  
\end{eqnarray}
so that the diagonal term arising from the 
pairs $(\gamma,\gamma)$ and $(\gamma,
{\bar \gamma})$ is 
\begin{equation} 
K_{\rm diag}(\tau) = K_{(\gamma,\gamma)}(\tau;\eta,\eta) +   
K_{(\gamma,{\bar \gamma})}(\tau;\eta,\eta) =  
2 \tau (1 + 3 {\rm e}^{- 2 a T}).  
\end{equation}
\par
Let us next consider the second order term. As before, 
it can be evaluated from the Sieber-Richter pair 
$(\gamma,\gamma')$ in Figure 1. We compute the average 
of $({\rm tr}\Delta_{\gamma}(\eta)) ({\rm tr} 
\Delta_{ \gamma'}(\eta))$ over the Brownian motion as  
\begin{eqnarray} 
& & \langle \langle \ ({\rm tr}\Delta_{\gamma}(\eta)) ({\rm tr}\Delta_{
\gamma'}(\eta)) \ \rangle \rangle \nonumber \\ 
& = & 
\int {\rm d}\omega_{L_1} 
{\rm d}\omega_{L_2} {\rm d}\omega_{E_1} 
\nonumber \\ & \times & 
{\rm tr}((\Delta_{E_1})^{-1} \Delta_{L_2} \Delta_{E_1} \Delta_{L_1}) 
{\rm tr}((\Delta_{E_1})^{-1} (\Delta_{L_2})^{-1} \Delta_{E_1} \Delta_{L_1})
\nonumber \\ & \times & g(\omega_{L_1};T_1|0,0,0)    
g(\omega_{L_2};T - T_1 - 2 t_1|0,0,0)    
g(\omega_{E_1};t_1|0,0,0)  
\nonumber \\ 
& = & 
- \frac{1}{2} + \frac{3}{2} {\rm e}^{- 2 a T + 4 a t_1} 
+ \frac{3}{2} {\rm e}^{- 2 a T + 2 a T_1 + 4 a t_1} + 
\frac{3}{2} {\rm e}^{- 2 a T_1}.  
\end{eqnarray}
Then we can evaluate the contribution to the form factor
\begin{eqnarray}
K_{\rm SR}(\tau) & = & 
\frac{4 \tau^2}{N_{\rm SR}} \left. \frac{\partial}{\partial t_1} 
\left\{ \int_0^{T - 2 t_1} {\rm d}T_1 \langle \langle \ 
({\rm tr}\Delta_{\gamma}(\eta)) ({\rm tr}\Delta_{
\gamma'}(\eta)) \ \rangle \rangle \right\} 
\right|_{t_1 = 0}
\nonumber \\ 
& = & 2 \tau^2 \left\{ 1 + (6 a T - 9) {\rm e}^{- 2 a T} \right\}.  
\end{eqnarray} 
Thus we obtain the semiclassical form factor up to the second order 
\begin{eqnarray}
\label{ksc2}
K_{\rm SC}(\tau) & = & K_{\rm diag}(\tau) +  K_{\rm SR}(\tau) 
\nonumber \\ 
& = &  2 \tau (1 + 3 {\rm e}^{- 2 a T}) 
+ 2 \tau^2 \left\{ 1 + (6 a T - 9) {\rm e}^{- 2 a T} \right\}.  
\end{eqnarray}
\par
A random matrix model of the GOE to GSE transition was already 
formulated in \cite{BWH,APWB}. However, as far as 
the authors know, an asymptotic formula to be compared with the 
above result (\ref{ksc2}) has not been worked out. Therefore it 
can be regarded as a conjecture for one of the open problems 
in random matrix theory.   
\par
The corresponding random matrix model can be formulated by using 
Dyson's p.d.f. (\ref{prm}). Here we need to suppose that the initial 
matrix $H_0$ is a GOE random matrix. Namely, the matrix elements 
of $H_0$ only have the $0$-th components and the p.d.f. of $H_0$ is 
\begin{equation}
P_{\rm GOE}(H_0) {\rm d}H_0 \propto {\rm e}^{- (1/2) {\rm Tr} (H_0)^2} 
{\rm d}H_0
\end{equation}
with
\begin{equation} 
{\rm d}H_0 = 
\prod_{j=1}^N {\rm d}(H_0)_{jj} \prod_{j<l}^N {\rm d}(H_0)_{jl}.
\end{equation}
It is well known that the form factor of the GOE eigenvalues 
is expanded as 
\begin{equation}
K_{\rm GOE}(\tau) = 2 \tau - 2 \tau^2 + \cdots.
\end{equation}
Considering the Kramers degeneracy, one modifies it into
\begin{equation}
{\tilde K}_{\rm GOE}(\tau) = 2 K_{\rm GOE}(2 \tau) = 
8 \tau - 16  \tau^2 + \cdots,
\end{equation}
which is in agreement with the corresponding case $a = 0$ 
of the semiclassical result (\ref{ksc2}). 
  
\section{Summary}
\setcounter{equation}{0}
\renewcommand{\theequation}{6.\arabic{equation}} 

In this paper, the parametric spectral correlation 
of a chaotic system with spin $1/2$ was studied. 
The parameter was chosen to be the strength of the 
effective field applied to the spin. Using the 
semiclassical periodic orbit theory for the orbital 
motion and simulating the spin dynamics by Brownian 
motion on a sphere, we evaluated the parameter-dependent 
spectral form factor $K_{\rm SC}(\tau)$. 
The $\tau$ expansion of $K_{\rm SC}(\tau)$ was found 
to be in agreement with the prediction of random 
matrix theory up to the third order. Moreover a 
crossover from a spinless system was investigated 
and the $\tau$ expansion of the corresponding 
form factor was calculated up to the second order.    

\section*{Acknowledgement}

One of the authors (T.N.) is grateful to Prof. Petr Braun, Dr. 
Sebastian M{\"u}ller, Dr. Stefan Heusler and Prof. Fritz Haake 
for valuable discussions.

\end{document}